\documentclass[notitlepage]{revtex4-1}

\newcommand{\eq}[1]{\begin{align}#1\end{align}}
\newcommand{\pert}[2]{
{}^{(#1)}\hspace{-0.5mm}#2
}
\newcommand{\Lxi}[1]{
\mathsterling_{{{}^{(#1)}}\xi}
}

\usepackage{amssymb, amsmath}
\usepackage{enumitem}
\usepackage{caption}
\usepackage{mathtools}
\usepackage{color}
\usepackage{natbib}
\usepackage{natbib}
\usepackage{hyperref}
\usepackage{enumitem}

\begin{document}

\title{The problem of ultracompact rotating gravastars}
\author{Mieszko Rutkowski}
\email{mieszko.rutkowski@doctoral.uj.edu.pl}
\author{Andrzej Rostworowski}
\affiliation{Institute of Theoretical Physics, Jagiellonian University, 30-348 Krak\'ow, Poland}

\begin{abstract}
A number of authors provided arguments that a rotating gravastar is a good candidate for a source of the Kerr metric. These arguments were based on the second order perturbation analysis.
In the following paper, we construct a perturbative solution of the rotating gravastar up to the third perturbation order and show that it cannot be continuously matched with the Kerr spacetime.
\end{abstract}

\maketitle

\section{Introduction}
\label{sect:1}
Gravastars, proposed by Mazur and Mottola \citep{MM04} as an alternative to black holes, have been studied extensively in the recent years (\citep{CFV05,C05,C07,CPCC08,CR08,PBCCN09,PBCCN10,CR16}). One of the issues concerning gravastars is to find a rotating gravastar solution. So far only perturbative versions of such a solution exist (\citep{U15,UYP16,Posada17}). These studies indicate that in the ultracompact limit (\citep{MM15}) the rotating gravastar can be a source of the Kerr metric (i.e.\ I, Love, Q numbers tend to those of Kerr in this limit). Similar perturbation-type sources (thin shells) of the Kerr metric  were studied earlier by e.g,\ \citep{C67,CI68,P86}. On the other hand, constructing perturbation sources of the Kerr metric have been criticised by \citet{K78}. 

In this work, we take perturbation approach to check if the matching of the gravastar with the Kerr spacetime survives at higher orders. It means that we want to construct a rotating analogue of \citep{MM15} with the Kerr spacetime outside. We use slightly different framework to (\citep{U15,UYP16,Posada17}) and instead of solving Einstein equations both for interior and exterior, we \textit{a priori} assume that an exterior solution is the Kerr metric. Then we seek for an interior solution and try to match it with the Kerr metric.

Most of the work on rotating gravastars was based on Hartle's structure equations \cite{H67} (see also \citep{Reina15,Mars20a,Mars20b}). Hartle's framework allows to study slowly rotating perfect fluid objects up to the second order in angular momentum. To go beyond the second order, we find it easier to follow \citet{AR17}, who provided a nonlinear extension of Regge-Wheeler and Zerilli formalisms. Formalism given by \cite{AR17} is dedicated to ($\Lambda$-) vacuum spacetimes and can be easily adapted to our needs. The difference between Hartle's framework and our approach is only on the level of ansatz on metric perturbation form and they are physically equivalent within the range of applicability of Hartle's framework. We find that \citep{AR17} provides a very powerful tool for dealing with nonlinear perturbations. Although in the present article we describe perturbation analysis only up to the third order, we solved Einstein equations up to the sixth order to calculate Kretschmann scalar and we think it's possible to go further if needed.

The paper is organised as follows: in Sections \ref{sect:2}, \ref{sect:3} and \ref{sect:4} we provide preliminaries, in Section \ref{sect:5} we discuss the matching, in Section \ref{sect:6} we expand the Kerr metric, in sections \ref{sect:7} and \ref{sect:8} we solve interior Einstein equations and try to match interior and exterior metrics and in Section \ref{sect:9} we summarise and discuss our calculations.

\section{Background solution}
\label{sect:2}
As a background, we take the ultracompact gravastar model \citep{MM15}. In static coordinates ($t$,$r$,$u$,$\varphi$), where $u=\cos\theta$, it's metric is given by:
\eq{
\bar g = f(r) dt^2 +\frac{1}{h(r)}dr^2 + r^2\left(\frac{du^2}{1-u^2}+(1-u^2) d\varphi^2\right)\, ,\label{g0stat}
}
where
\eq{\label{g0}
 f(r)=& \left\{
        \begin{array}{ll}
            \frac{1}{4}\left(1-\frac{r^2}{4M^2}\right) & \quad r \leq R \, ,\\
            1-\frac{2M}{r} & \quad r > R\, ,
        \end{array}
    \right.\\
    h(r)=& \left\{
        \begin{array}{ll}
            1-\frac{r^2}{4M^2} & \quad r \leq R \, ,\\
            1-\frac{2M}{r} & \quad r > R\, .
        \end{array}
    \right.
}
An induced metric is continuous across the (null) matching surface $r=2M$. There is a nonzero stress-energy tensor induced on this shell, see \cite{MM15} for the details. The exterior metric is a solution to vacuum Einstein equations and the interior metric is a solution to Einstein equations with a cosmological constant $\Lambda = \frac{3}{4 M^2}$. Both interior and exterior metrics are singular at $r=2M$. To keep them regular, also in higher perturbation orders, we use Eddington--Finkelstein (EF) coordinates ($v$,$r$,$u$,$\varphi$). Interior metric in EF coordinates reads:
\eq{
\bar g = \frac{1}{4}\left(1-\frac{r^2}{4M^2}\right) dv^2 + dr d v + r^2\left(\frac{du^2}{1-u^2}+(1-u^2) d\varphi^2\right)\, .
}
and exterior metric in EF coordinates reads:
\eq{
\bar g =\left(1-\frac{2M}{r}\right)dv^2 + 2dr d v + r^2\left(\frac{du^2}{1-u^2}+(1-u^2) d\varphi^2\right)\, .
}

\section{Polar expansion}
\label{sect:3}
In a spherically symmetric background, in 3+1 dimensions, vector and tensor components split into two sectors: polar and axial (for the details see e.g.\ \citep{RW57,Z70a,Z70b,HPN99,SM00}). Symmetric tensors have 7 polar and 3 axial components. Below we list the expansion of the components of symmetric tensors in axial symmetry ($P_\ell$ denotes the $\ell$-th Legendre polynomial). 

The symmetric tensor, polar sector:
\eq{
&S_{ab}(r,u)=\sum\limits_{0\leq \ell} {S_\ell}_{ab}(r) P_\ell(u)\,, \quad a,b=v,r\, ,\label{sp}\\
&S_{au}(r,u)=-\sum\limits_{1\leq \ell} {S_\ell}_{a u}(r)\partial_u P_\ell(u)\,, \quad a=v,r\, ,\\
&\frac{1}{2}\left((1-u^2)S_{uu}(r,u)+\frac{S_{\varphi\varphi}(r,u)}{(1-u^2)}\right)=\sum\limits_{0\leq \ell} {S_\ell}_{+}(r) P_\ell(u)\, ,\\
&\frac{1}{2}\left((1-u^2)S_{uu}(r,u)-\frac{S_{\varphi\varphi}(r,u)}{(1-u^2)}\right)= \nonumber \\
&=\sum\limits_{2\leq \ell} {S_\ell}_{-}(r) (-\ell(\ell+1)P_\ell(u)+2u\partial_u P_\ell(u))\, .
}
The symmetric tensor, axial sector:
\eq{
&S_{a\varphi}(r,u)=\sum\limits_{1\leq \ell} {S_\ell}_{a\varphi}(r) (-1+u^2) \partial_u P_\ell(u)\,, \quad a=v,r\, ,\\
&S_{u\varphi}(r,u)=\sum\limits_{2\leq \ell} {S_\ell}_{u\varphi}(r) \left(\ell (\ell+1)  P_\ell(u)-2 u \partial_u P_\ell(u)\right) \, .\label{sa}
}



\section{Metric perturbations}
\label{sect:4}
We assume that there exists an exact, stationary and axially symmetric solution to Einstein equations, which we expand into series in a parameter $a$ (which will be an angular momentum per unit mass of a an exterior metric) around the static metric \eqref{g0}:
\eq{
g_{\mu\nu} = \bar g_{\mu\nu} + \sum\limits_{i=1}^\infty \frac{a^i}{i!}{}^{(i)}h_{\mu\nu}
}

After perturbation expansion we polar--expand metric perturbations according to \eqref{sp} - \eqref{sa}. Thus, apart from the perturbation index $i$, all perturbations gain an index $\ell$ corresponding to the $\ell$-th Legendre polynomial.

For axial perturbations we take:
\eq{
{}^{(i)}h_\ell=\begin{pmatrix}
    0&0 & 0 &\pert{i}h_{\ell\, v\varphi} (r) (-1+u^2) \partial_u P_\ell(u) \\
    0& 0 & 0 & \pert{i}h_{\ell\, r\varphi} (r) (-1+u^2) \partial_u P_\ell(u) \\
    0&0& 0 & 0 \\
    \pert{i}h_{\ell\, v\varphi} (r) (-1+u^2) \partial_u P_\ell(u)&\pert{i}h_{\ell\, r\varphi} (r) (-1+u^2) \partial_u P_\ell(u)& 0 & 0 
\end{pmatrix}\, .
}
Using  the gauge freedom, we set ${}^{(i)}h_{\ell\, u\varphi} (r) =0$, what corresponds to the Regge-Wheeler (RW) gauge.

For the polar perturbations we take:
\eq{
{}^{(i)}h_\ell=\begin{pmatrix}
    {}^{(i)}h_{\ell\, vv} (r)P_\ell(u)&{}^{(i)}h_{\ell\, vr} (r)P_\ell(u) & 0 &0 \\
    {}^{(i)}h_{\ell\, vr} (r)P_\ell(u)& {}^{(i)}h_{\ell\, rr} (r)P_\ell(u) & 0 & 0 \\
    0&0& \pert{i}h_{\ell\, +} (r)\frac{P_\ell(u)}{1-u^2}& 0 \\
    0&0& 0 & \pert{i}h_{\ell\, +} (r) \left(1-u^2\right)P_\ell(u)
\end{pmatrix}\, .
}
Using the gauge freedom, we set ${}^{(i)}h_{\ell\, ru} ={}^{(i)}h_{\ell\, vu} ={}^{(i)}h_{\ell\, -} =0$, what also corresponds to the RW gauge. Note that in \cite{H67} there are no ${}^{(i)}h_{\ell\, vr} $ and ${}^{(i)}h_{\ell\, r\varphi} $ coefficients in the metric ansatz. This fact arises from the fact that Hartle uses static coordinates. For EF coordinates in the background both ${}^{(i)}h_{\ell\, vr} $ and ${}^{(i)}h_{\ell\, r\varphi} $ turn out to be nonzero in most cases. 

In the interior, we solve perturbation Einstein equations with a cosmological constant $\Lambda = \frac{3}{4 M^2}$. For a given order $i$ and a given multipole $\ell$, they have the following form:
\eq{\label{ee}
\delta G_{\ell\, \mu\nu}= {}^{(i)}S_{\ell\, \mu\nu}\, ,
}
where $\delta G_{\ell\, \mu\nu}$ denotes the components of the Einstein tensor expansion. $ \pert{i} S_{\mu\nu}$  denotes a source for the $i$-th order Einstein equations consisting of metric perturbations of orders lower that $i$. We provide an explicit form of equations \eqref{ee} in the Appendix \ref{appb}.


\section{Matching interior with exterior}
\label{sect:5}
We match the exterior metric with the interior metric on a three-dimensional hypersurface located at $r^{\pm}=r_b^\pm$, where ``+'' and ``-'' stand for exterior and interior, respectively. From the first Israel junction condition (\citep{Israel66, BI91}) we demand continuity of the induced metric at the matching hypersurface:
\eq
{
[[\mathfrak{g}^\pm_{ab}]]=0\, ,\label{is1}
}
where $[[E]]=E^{+}(r_b^+)-E^{-}(r_b^-)$. Following \citep{UYP16}, we introduce intrinsic coordinates on the three-dimensional hypersurface: $y^a = (V,U,{\Phi})$. Then we express interior and exterior coordinates ${x}^{\pm\, \mu}$ on a hypersurface in terms of $y^a$:
\eq{
x^{- \mu} \big |_{r_b^-} &= \left(A^-\, V,r_b^-(U) , F^-(U) , \Phi\right)\, ,\\
x^{+ \mu} \big |_{r_b^+} &= \left(A^+\, V,r_b^+(U) , F^+(U) , \Phi \right)\, ,
}
where $r_b^{\pm}(U) =   2M+\frac{a^2}{M^2}\eta^{\pm}(U)+\mathcal{O}(a^4)$, $F^{\pm}(U) = U+\frac{a^2}{M^2}\lambda^{\pm}(U)+\mathcal{O}(a^4)$. We expand $\eta^{\pm}$ into $\eta^\pm(U)=\eta_0^\pm+\eta_2^\pm P_2(U) $. 

The metric induced on this hypersurface is given by:
\eq{
\mathfrak{g}^\pm_{ab}=\left(
\begin{array}{ccc}
 \left(A^{\pm}\right)^2 g^\pm_{vv} & A^{\pm} g^\pm_{vr}{r^\pm_b}'(U)+ A^{\pm} g^\pm_{vu} {F^{\pm}}'(U)  & A^{\pm} g^\pm_{v\varphi} \\
 A^{\pm} g^\pm_{vr}{r^\pm_b}'(U)+ A^{\pm} g^\pm_{vu} {F^{\pm}}'(U)  &\left({F^\pm}'(U)\right)^2 g^\pm_{uu}+\left({r^\pm_b}'(U)\right)^2 g^\pm_{rr} + 2{F^\pm}'(U) {r^\pm_b}'(U) g^\pm_{ru} & {F^{\pm}}'(U)g^\pm_{u\varphi}+{r^\pm_b}'(U) g^\pm_{r\varphi} \\
A^{\pm} g^\pm_{v\varphi}  & {F^{\pm}}'(U)g^\pm_{u\varphi}+{r^\pm_b}'(U) g^\pm_{r\varphi} & g^\pm_{\varphi\varphi} \\
\end{array}
\right)\, .
}
Using the freedom in a choice of coordinates $V,U,\Phi$, we set $F^{+}(U) = U$ and $A^+=1$ (see e.g. \cite{U15}). For simplicity, we denote $A^-=A$.

The location of the matching hypersurface is not known \textit{a priori} and $\eta^\pm(U)$ and $\lambda^-(U)$ are unknown functions that need to be found. Our procedure of matching interior and exterior metrics for a given perturbation order is the following:

\begin{enumerate}
\item We solve perturbation Einstein equations for the interior. These solutions contain two constans per $\ell$ in every perturbation order, but most of these constants need to be set to zero to keep Kretschmann scalar regular at $r=0$ and $r=2M$. However, this is not straightforward to apply, because in our case singularities of the Kretschmann scalar occur in higher perturbation orders than the singularities of the metric itself (in the opposition to the exterior case, e.g.\ \citet{RPE19}). Therefore, to settle constants in the third order, we solved Einstein equations up to the sixth perturbation order to study behaviour of the Kretschmann scalar. Since these expressions are too long to be listed in this paper, we make them available in the Mathematica notebook \citep{KretschmannNotebook}.
\item We act with the general gauge transformation on the interior metric, and then we solve matching conditions \eqref{is1} for constants arising from Einstein equations, for $\eta^\pm(U),\,  \lambda(U) ,$ and for gauge components. Finding a proper gauge is a part of the matching problem and using the result of \citet{B97}, we are able to control the impact of the gauge from the lower perturbation order on the metric functions in the higher perturbation order.
\item If the matching is successful, we go to the higher perturbation order.
\end{enumerate}

The second junction condition tells about the energy content of the matching hypersurface - already in the background solution there is a thin shell located at $r=2M$ (since this is a null hypersurface, second junction condition needs to be modified, see \citep{BI91}, \citep{MM15} for the details). However, in the next sections we show that even the first junction condition is not possible to fulfil, therefore we don't find it necessary to discuss second junction condition at all.

\section{Kerr metric expansion}
\label{sect:6}
As an exterior metric, we take the Kerr solution. In the advanced EF coordinates it reads:
\eq{
ds^2 =&-  \left(1-\frac{2 M r}{a^2 u^2+r^2}\right) dv^2 +2 dv dr + \frac{a^2 u^2+r^2}{1-u^2} du^2 + \left(1-u^2\right) \left(\frac{2 a^2 M r \left(1-u^2\right)}{a^2 u^2+r^2}+a^2+r^2\right) d\varphi^2 +\nonumber\\
&+  \frac{4 a M r \left(1-u^2\right)}{a^2 u^2+r^2} dv d\varphi + 2 a \left(1-u^2\right) dr d\varphi\, .\label{kerrMetric}
}
Since we solve the interior equations in RW gauge, we prefer to use the Kerr metric in RW gauge as well. To do this, we expand \eqref{kerrMetric} into series in $a$ up to the 3rd order, and then act with the gauge transformations \eqref{gg1}-\eqref{gg3} to move to the RW gauge. Finally, we obtain:
\eq{
ds^2 =&-  \left(\left(1-\frac{2 M}{r}\right)-\frac{a^2 M \left(u^2 \left(6 M^2-M r-3 r^2\right)-2 M^2+M r+r^2\right)}{r^5}\right) dv^2  +\nonumber\\
&+  \left(\frac{2 a^2 M \left(1-3 u^2\right)}{r^3}\right)dr^2+ \left(\frac{r^2}{1-u^2}+\frac{a^2 M \left(3 u^2-1\right) (2 M+r)}{r^2 \left(u^2-1\right)}\right)du^2\nonumber\\
&+\left(r^2 \left(1-u^2\right)+\frac{a^2 M \left(u^2-1\right) \left(3 u^2-1\right) (2 M+r)}{r^2}\right) d\varphi^2 +\nonumber\\
&+2\left(1+\frac{ a^2 M \left(3 u^2-1\right) (M+r)}{r^4}\right) dv dr +  2\left(\frac{a^3 M \left(1-u^2\right) \left(5 u^2-1\right) (9 M+5 r)}{5 r^4}\right)dr d\varphi + \nonumber\\
&+ 2 \left(\frac{2 a M \left(1-u^2\right)}{r} -\frac{a^3 M \left(u^2-1\right) \left(M^2 \left(6 u^2-2\right)+M \left(r-5 r u^2\right)+r^2 \left(1-5 u^2\right)\right)}{r^5}\right) dv d\varphi\, +\mathcal{O}(a^4),\label{kerrMetricRW}
}
For simplicity, we omit ``+'' and ``-'' coordinate superscripts and use them only when it's necessary to differentiate the interior from the exterior. We expand \eqref{kerrMetricRW} into series in $a$. Below we list nonzero components of this expansion after the polar decomposition.
\eq{
\begin{split}
{}^{(1)}h^ {+}_{ 1,\, v\varphi}&=           -\frac{2 M}{r}                     \, ,\\
{}^{(2)}h^ {+}_{ 0,\, vv}&=      \frac{4 M^2}{3 r^4}                           \, ,\\
{}^{(2)}h^ {+}_{ 2,\, vv}&=          \frac{4 M \left(6 M^2-M r-3 r^2\right)}{3 r^5}                      \, ,\\
{}^{(2)}h^ {+}_{ 2,\, vr}&=         \frac{4 M (M+r)}{r^4}                      \, ,\\
{}^{(2)}h^ {+}_{ 2,\, rr}&=       -\frac{8 M}{r^3}                 \, ,
\end{split}
\begin{split}
{}^{(2)}h^ {+}_{ 2,\, +}&=            -\frac{4 M (2 M+r)}{r^2}                    \, ,\\
{}^{(3)}h^ {+}_{ 1,\, v\varphi}&=          \frac{24 M^3}{5 r^5}                      \, ,\\
{}^{(3)}h^ {+}_{ 3,\, v\varphi}&=                     \frac{4 M \left(-6 M^2+5 M r+5 r^2\right)}{5 r^5}                   \, ,\\
{}^{(3)}h^ {+}_{ 3,\, r\varphi}&=                          -\frac{4 M (9 M+5 r)}{5 r^4}              \, .
\end{split}
}

\section{Interior solution}
\label{sect:7}

\subsection{The first order}
\subsubsection{Axial $\ell=1$}
For $\ell=1$ there is no $h_{u\varphi}$ component and we can use the remaining gauge freedom to set $\pert{1}h^-_{1\,r\varphi}=0$. Linearized Einstein equation are homogeneous \eqref{eq:axial1}-\eqref{eq:axial3} and yield:
\eq{
\pert{1}h^ {-}_{1\, v \varphi }= \Omega_{11} r^2 + \frac{\Pi_{11}}{r}
}
where $\Omega_{11}$ and $\Pi_{11}$ are arbitrary constants. We set $\Pi_{11}=0$ to make Kretchmann scalar regular at r=0, therefore we are left with $\pert{1}h^-_{1\, v \varphi }= \Omega_{11} r^2 $. It turns out that this solution is a pure gauge, but we will discuss it later.

\subsection{The second order}
\subsubsection{Polar $\ell =0$}
For $\ell=0$ there are no $h_-,\, h_{vu},\, h_{ru}$ components in the polar decomposition and we have an additional gauge freedom, which we use to set $\pert{2}h^ {-}_{0\, vr},\,\pert{2}h^ {-}_{0\, +}$ to zero. The only nonzero variables left are $\pert{2}h^ {-}_{0\,vv}$ and $\pert{2}h^ {-}_{0\,rr}$.

Solution to Einstein equations \eqref{eq:polar1}-\eqref{eq:polar6} with $\ell=0$ and with sources \eqref{eq:s201}-\eqref{eq:s202} reads:
\eq{
\pert{2}h^-_{0\,vv} &=\frac{4 r^2 \Omega_{11} ^2}{3}-\frac{c_{20} \left(r^2-4 M^2\right)}{64 M^4}+\frac{d_{20}}{r} \, ,\\
\pert{2}h^-_{0\,rr} &=     \frac{c_{20}}{r^2-4 M^2}\, . 
}
where $c_{20}$ and $d_{20}$ are arbitrary constants. This solution is singular at $r=0$ and $r=2M$. To avoid singularity in Kretschmann scalar at r=0, we set $d_{20}=0$. Singularity at $r=2M$ can be removed using a gauge transformation ($\pert{2}\xi_{0\, v}=\frac{c_{20} \left(\left(r^2-4 M^2\right) \tanh ^{-1}\left(\frac{r}{2 M}\right)+2 M r\right)}{64 M^3}$, $\pert{2}\xi_{0\, r}= \frac{c_{20} \tanh ^{-1}\left(\frac{r}{2 M}\right)}{8 M}$), what yields:
\eq{
\pert{2}h^ {-}_{0\, vv}&=  \frac{4 r^2   \Omega_{11} ^2}{3}+\frac{c_{20}}{16 M^2} \, ,\\
\pert{2}h^ {-}_{0\, vr}&=0 \, ,\\
\pert{2}h^ {-}_{0\, rr}&=0\, ,\\
\pert{2}h^ {-}_{0\, +}&= \frac{c_{20} r^2}{4 M^2} \, .
}

\subsubsection{Polar $\ell =2$}
Solution to Einstein equations \eqref{eq:polar1}-\eqref{eq:polar6} with $\ell=2$ and with sources \eqref{eq:s221}-\eqref{eq:s222} reads:
\eq{
\pert{2}h^-_{2\, vv}=&     \frac{ \left(r^2-4 M^2\right)^2}{128 M^4}  \pert{2}h^-_{2\, rr}      - \frac{4}{3}  r^2   \Omega_{11} ^2   \,  ,\\
\pert{2}h^-_{2\, vr}=&  -\frac{1}{4} \left(1-\frac{r^2}{4M^2}\right)  \pert{2}h^-_{2\, rr}   \, ,\\
\pert{2}h^-_{2\, rr}=&   \frac{c_{22}}{16 M^4 r^3}+\frac{d_{22} \left(3 \left(r^2-4 M^2\right)^2 \coth ^{-1}\left(\frac{2 M}{r}\right)+2M r(5r^2-12M^2)\right)}{32 M^3 r^3 \left(r^2-4 M^2\right)^2}   \, ,\\
\pert{2}h^-_{2\, +}=&   \frac{c_{22} \left(4 M^2+r^2\right)}{128 M^6 r}+\frac{d_{22} \left(3 M \left(4 M^2+r^2\right) \coth ^{-1}\left(\frac{2 M}{r}\right)-2r \left(3 M^2 +r^2\right)\right)}{256 M^6 r} \, ,
}
where $c_{22}$ and $d_{22}$ are arbitrary constants. To avoid singularity in the Kretschmann scalar at $r=0$ and $r=2M$ we need to set $c_{22}=0$, $d_{22}=0$, what yields:
\eq{
\pert{2}h^-_{2\, vv}&=   - \frac{4}{3}  r^2   \Omega_{11} ^2 ,  ,\\
\pert{2}h^-_{2\, vr}&=       0   \, ,\\
\pert{2}h^-_{2\, rr}&=        0    \, ,\\
\pert{2}h^-_{2\, +}&=0\,.  
}

\subsection{The third order}
\subsubsection{Axial $\ell=1$}
The solution to Einstein equations \eqref{eq:axial1}--\eqref{eq:axial3} with $\ell=1$ reads:
\eq{
\pert{3}h^ {-}_{1\, v\varphi}= \Omega _{31}r^2+\frac{\Pi _{31}}{r}\, .\label{sol31}
}
To keep Kretschmann scalar regular at r=0, we set $\Pi_{31}=0$.
 
\subsubsection{Axial $\ell=3$}
Solution to Einstein equations \eqref{eq:axial1}-\eqref{eq:axial3} with $\ell=3$ reads:
\eq{
\pert{3}h^ {-}_{3\,v\varphi}& =\frac{\left(r^2-4 M^2\right)}{r^3}  \Pi _{33}  +\frac{ \left(-120 M^4 r+20 M^2 r^3+60 \left(4 M^5-M^3 r^2\right) \coth ^{-1}\left(\frac{2 M}{r}\right)+r^5\right)}{3 r^3}\Omega _{33}  \, ,\label{sol33a}\\
\pert{3}h^ {-}_{3\,r\varphi}& = \frac{8 M^2 }{r^3}\Pi _{33}+\frac{8 M^2 \left(\frac{r \left(-120 M^4+20 M^2 r^2+r^4\right)}{r^2-4 M^2}-60 M^3 \coth ^{-1}\left(\frac{2 M}{r}\right)\right)}{3 r^3}  \Omega _{33} \, ,\label{sol33b}
}
where $\Omega_{33}$ and $\Pi_{33}$ are arbitrary constants. Singularities at $r=0$ and $r=2M$ lead to the singularity in the Kretschmann scalar, therefore $\Omega_{33}=0$, $\Pi_{33}=0$.

\section{Matching}
\label{sect:8}

\subsubsection{First order}
Before matching, we act with the general gauge transformation on the interior metric. Although we consider stationary metrics, we take gauge vectors that depend on $v$ coordinate. It might happen, that acting with gauge vectors depending on $v$ explicitly, we obtain metric independent of $v$ (we discuss such a case in Section \ref{sect:9}). From the matching conditions \eqref{is1} we have:
\eq{
\frac{\pert{1}h^+_{1\, v \varphi }(2M)}{A}-\pert{1}h^-_{1\, v \varphi }(2M)=-\partial_v\pert{1}\xi_{1\,\varphi}(v,2M)\, ,\label{isl1a}
}
To keep transformed metric $v$-independent, we use \eqref{ga1} and \eqref{ga2} and obtain a condition:
\eq{
\pert{1}\xi_{1\, \varphi }= q_{11} v r^2 + \pert{1}\gamma_{1\, \varphi}(r)\, ,
}
where $q_{11}$ is an arbitrary constant and $\gamma_{1}$ is an arbitrary function of $r$. From \eqref{isl1a} we obtain:
\eq{
\Omega_{11} = -\frac{1}{4 A M^2}+q_{11}\, .\label{i11a}
}

\subsubsection{Second order}
We act with the most general second order gauge transformation \eqref{gg1}-\eqref{gg2} on the interior metric. To keep transformed metric $v$-indepedent, we use \eqref{gp1}-\eqref{gp6} and obtain conditions:
\eq{
\pert{2}\xi_{0\, v} = -4M^2 f q_{20} v + \pert{2}\gamma_{0\, v}(r)\, ,\\
\pert{2}\xi_{0\, r} = 8 M^2 q_{20} v + \pert{2}\gamma_{0\, r}(r)\, ,\\
\pert{2}\xi_{2\, v} =  \pert{2}\gamma_{2\, v}(r)\, ,\\
\pert{2}\xi_{2\, r} =  \pert{2}\gamma_{2\, r}(r)\, ,\\
\pert{2}\xi_{2\, u} = \pert{2}\gamma_{2\, u}(r)\, ,
}
where $q_{20}$ is an arbitrary constant and $\pert{i}\gamma_{\ell\, \mu}$ are functions of $r$. 

Matching conditions \eqref{is1} yield:
\eq{
\pert{2}h^+_{0\, vv}(2M)-A^2\pert{2}h^-_{0\, vv}(2M)& =     \frac{A^2 \eta_0^-+2 \eta_0^+}{2 M^3} +\frac{16}{3} A^2 M^2 q_{11} \left(q_{11}-2 \Omega _{11}\right) +\frac{A^2}{2 M} \pert{2}\gamma_{0\, v}(2 M)\, ,\label{i20}\\
\pert{2}h^+_{2\, vv}(2M)-A^2\pert{2}h^-_{2\, vv}(2M)& = \frac{A^2\eta_2^-+2 \eta_2^+}{2 M^3}    -\frac{16}{3} A^2 M^2 q_{11} \left(q_{11}-2 \Omega _{11}\right)     +\frac{A^2}{2 M} \pert{2}\gamma_{2\, v}(2M) \, ,\\
2\eta_2^+-  \eta_2^- A&=A M^2 \pert{2}\gamma _{2\,v}(2 M) \, ,\\
[[\pert{2}h_{0\, +}(2M)]]& = - \frac{8 (\eta_0^+ - \eta_0^-)}{M} + 8 \lambda '(U) +8 M\pert{2}\gamma_{0\, v}(2 M)\, ,\\
[[\pert{2}h_{2\, +}(2M)]]& = -\frac{8 ( \eta_2^+- \eta_2^-)}{M} +8 M \pert{2}\gamma_{2\, v}(2 M)-6 \pert{2}\gamma_{2\, u}(2 M)\, ,\\
[[\pert{2}h_{2\, -}(2M) ]]&=\pert{2}\gamma_{2\,u}(2 M) +\frac{16 U \lambda (U) + 8 \left(1-U^2\right) \lambda '(U)}{3 \left(U^2-1\right)^2}\, .\label{i22}
}
After plugging solutions to perturbation equations into \eqref{i20}-\eqref{i22}, we obtain:
\eq{
\eta_0^-&-= -  M^2\pert{2}\gamma_{0\,v}(2 M)-\frac{4 M U}{3} \lambda_1-\frac{ M}{8}c_{20}-\frac{M}{6} -\frac{1}{4} M \left(3 U^2-1\right) \pert{2}\gamma_{2\,u}(2 M)\,, \\
\eta_2^-&=  -\frac{M}{3} - M^2\pert{2}\gamma_{2\,v}(2 M)+\frac{1}{2} M \pert{2}\gamma_{2\,u}(2 M) \,, \\
\eta_0^+&=  -\frac{M}{6} +\frac{2  \lambda_1M U}{3}+\frac{1}{8} M \left(3 U^2-1\right) \pert{2}\gamma_{2\,u}(2 M)  \,, \\
\eta_2^+&=     \frac{M}{6}  -\frac{1}{4} M \pert{2}\gamma_{2\,u}(2 M)  \,, \\
A &= -1 \, ,\\
\lambda(U)&=  \lambda_1\left(U^2-1\right)+\frac{3}{8} U \left(U^2-1\right) \pert{2}\gamma_{2\,u}(2 M) \, .
}
where $\lambda_1$ is an arbitrary constant. To keep $\eta_0^-$ independent of $U$, we have to set $\lambda_1 = 0$ and $\pert{2}\gamma_{2\,u}(2 M)=0$, what leads to:
 \eq{
\eta_0^-&-= -  M^2\pert{2}\gamma_{0\,v}(2 M)-\frac{ M}{8}c_{20}-\frac{M}{6} \,, \label{i22a}\\
\eta_2^-&=  -\frac{M}{3} - M^2\pert{2}\gamma_{2\,v}(2 M)\,, \\
\eta_0^+&=  -\frac{M}{6}\,, \\
\eta_2^+&=     \frac{M}{6}   \,, \\
A &= -1 \, ,\\
\lambda(U)&=0 \, ,\\
\pert{2}\gamma_{2\,u}(2 M)&=0\, .\label{i22b}
}

\subsubsection{Third order}
Again, we act with the most general third order gauge transformation \eqref{gg1}-\eqref{gg3} on the interior metric. To keep transformed metric $v$-indepedent, we use \eqref{ga1}-\eqref{ga3} and obtain conditions:
\eq{
\pert{3}\xi_{1\,\varphi} =  q_{31} r^2 v + \pert{3}\gamma_{1\, \varphi}(r)\, ,\\
\pert{3}\xi_{3\,\varphi} =   \pert{3}\gamma_{3\, \varphi}(r)\, ,
}
where $q_{31}$ is an arbitrary constant and $\pert{i}\gamma_{\ell\, \mu}$ are functions of $r$. Using \eqref{i11a} and \eqref{i22a}-\eqref{i22b}, third order matching conditions \eqref{is1} yield:
\eq{
\pert{3}h_{1\,v\varphi}(2M)-A\pert{3}h_{1\,v\varphi}(2M)=& \frac{3(5 c_{20}+8)}{20 M^2}+3 c_{20} q_{11}-192 M^4 q_{11} q_{20}+M^2 4 \left( q_{31}-12 q_{20}\right)\, \label{i30},\\
\pert{3}h_{3\,v\varphi}(2M)-A\pert{3}h_{3\,v\varphi}(2M)=&\frac{3}{10 M^2}\, , \label{i31}\\
5M^2\pert{3}\xi _{3,\varphi }(2 M) =&6 \pert{2}\gamma_{2\, r}(2M) \left(4 M^2 q_{11}+1\right)+ 2 \left(3 M \pert{2}\gamma_{2\, v}(2M)+1\right) \left(M \pert{1}\gamma_{1\, \varphi}'(2M)-\pert{1}\gamma_{1\, \varphi}(2M)\right)\, .\label{i32}
}
Condition \eqref{i32} can be fulfilled just by setting all the gauge components to zero. Setting $\xi_{2\, u}=0$ and plugging \eqref{sol31}-\eqref{sol33b} into \eqref{i30}, we obtain:
\eq{
\Omega_{31}=\frac{9}{80 M^4}+q_{31}+\frac{3 \left(4 M^2 q_{11}+1\right) \left(c_{20}-64 M^4 q_{20}\right)}{16 M^4}\, .
}
However, \eqref{i31} does not have any free parameters and it cannot be fulfilled (we obtain contradiction $-\frac{3}{10M^2}=0$) . That makes impossible to match interior with exterior in the third order.

\section{Discussion and summary}
\label{sect:9}
Although we found the matching impossible, it is interesting to know what is the interior solution we obtained. The regular interior solution up to the third order reads:
\eq{
ds^2 =
\left(
\begin{array}{cccc}
-\frac{1}{4} \left(1-\frac{r^2}{4M^2}\right)+ a^2 \left(\frac{c_{20}}{32 M^2}+r^2 \left(1-u^2\right) \Omega _{11}^2\right) & \frac{1}{2} & 0 &\frac{1}{6} a r^2 \left(u^2-1\right) \left(6 \Omega _{11}+a^2 \Omega _{31}\right)\\
 \frac{1}{2} & 0 & 0 & 0 \\
 0 & 0 & \frac{r^2}{1-u^2}+\frac{a^2 c_{20} r^2}{8 M^2 \left(1-u^2\right)} & 0 \\
 \frac{1}{6} a r^2 \left(u^2-1\right) \left(6 \Omega _{11}+a^2 \Omega _{31}\right)& 0 & 0 & r^2 \left(1-u^2\right)+\frac{a^2 c_{20} r^2 \left(1-u^2\right)}{8 M^2}\\
\end{array}
\right)
\, .\label{full3}
}
It turns out that this is an exact solution to Einstein equations -- a gauge--transformed de Sitter space. To see this, let's take the gauge vector with components:
\eq{
\pert{1}\xi_1 =&\left(0,0,0,r^2 \Omega _{11} v\right)\, ,\\
\pert{2}\xi_0 =&\left( -\frac{c_{20} r}{16 M^2}+\frac{c_{20}  \left(r^2-4 M^2\right)}{128 M^4}v, \frac{c_{20} v}{16 M^2},0,0 \right)\, ,\\
\pert{2}\xi_2 =&\left(0,0,0,0\right)\, ,\\
\pert{3}\xi_1 =&\left(0,0,0,  (r^2 \Omega _{13}-\frac{3 c_{20} r^2 \Omega _{11}}{8 M^2}) v  \right)\, ,\\
\pert{3}\xi_3 =&\left(0,0,0,0 \right)\, .
}
Acting with those vectors on \eqref{full3} (using formulas \eqref{gg1}-\eqref{gg3}), we obtain
\eq{
ds^2=
\left(
\begin{array}{cccc}
 - \frac{1}{4} \left(1-\frac{r^2}{4M^2}\right) & \frac{1}{2} & 0 & 0 \\
 \frac{1}{2} & 0 & 0 & 0 \\
 0 & 0 & \frac{r^2}{1-u^2} & 0 \\
 0 & 0 & 0 & -r^2 \left(u^2-1\right) \\
\end{array}
\right)\, ,
}
what is exactly the background de Sitter metric, so all perturbations we obtained are a pure gauge. 
Therefore, from our calculations it follows that one cannot match a regular de Sitter vacuum with the Kerr metric, at least when in the limit $a\rightarrow 0$ one has the ultracompact gravastar solution. One can ask, if allowing for a change in the background density does not affect this result, but the answer is no. We repeated the calculation allowing for the perturbations of density and pressure (within the equation of state $p=-\rho$), but they do not change the conclusions. 

To sum up, we made an attempt to match the ultracompact rotating gravastar with the Kerr metric using the nonlinear perturbation theory. Although the matching can be performed up to the second order, in the third order it is is no longer possible, therefore the rotating gravastar in the discussed form is not a good candidate for the source of the Kerr metric. What's more, the interior of the ultracompact rotating gravastar is just the de Sitter metric. Since some of the proposed sources of the Kerr metric are based on the second perturbation order calculations, we find it necessary to check if these results survive at the higher perturbation orders.

\begin{acknowledgements}
This research was supported by the Polish National Science Centre grant no. 2017/26/A/ST2/00530. We wish to thank Prof. Pawe\l\ Mazur for a discussion.
\end{acknowledgements}

\appendix
\section{Einstein equations}\label{appb}
Einstein equations \eqref{ee} of order $i$ divide into two parts: the homogeneous part $\delta G_{\ell\, \mu\nu}$ consisting of metric perturbations of order $i$ and sources ${}^{(i)}S_{\ell\, \mu\nu}\,$ consisting of metric perturbations of orders $j$ ($j<i$). These equations needs to be solved order by order: after solving Einstein equations up to order $i$ one can construct explicit form of $i+1$ order source.

\subsubsection{Homogeneous part}

For the axial sector in the RW gauge, there are two nonzero variables: ${^{(i)}}h_{\ell v\varphi}$ and ${^{(i)}}h_{\ell r\varphi}$ (for simplicity, we denote ${}^{(i)}h_{\ell\, \mu\nu} (r)= h_{\mu\nu}$). Homogeneous parts of Einstein equations read:
\eq{
2 i! r^2 \left(\delta G\right)_{\ell\, v\varphi} &=h_{v\varphi} \left(2 f+\ell(\ell+1)-2\right)h_{v\varphi} -r^2 f h_{v\varphi} ''\, ,\label{eq:axial1}\\
2 i! r^2\left(\delta G\right)_{\ell\, r\varphi} &= 2 r^2 h_{v\varphi}'' -4 h_{v\varphi} +\left(\ell(\ell+1)-2\right) h_{r\varphi}  \, , \label{eq:axial2}\\
2 i!\left(\delta G\right)_{\ell\, u\varphi} &=f h_{r\varphi}'+2 h_{v\varphi}'+f' h_{r\varphi}\, .\label{eq:axial3}
}

For the polar sector in the RW gauge, there are four nonzero variables: ${^{(i)}}h_{\ell vv}$, ${^{(i)}}h_{\ell vr}$, ${^{(i)}}h_{\ell rr}$, ${^{(i)}}h_{\ell +}$ (for simplicity, we denote ${}^{(i)}h_{\ell\, \mu\nu} (r)= h_{\mu\nu}$). Homogeneous parts of Einstein equations read:
\eq{
 8i! r^4\left(\delta G\right)_{\ell\, vv} =&2 f^3 r^3 h_{rr}'+8 f^2 r^3 h_{vr}'-2 f^2 r^2 h_+''+4 f r^2 \left(2 r f'+2 f+\ell(\ell+1)\right)h_{vr} +\nonumber\\
&+f  \left(2 r f'+\ell(\ell+1)-2\right)h_++f r \left(2 f-r f'\right)h_+' +f^2 r^2 \left(4 r f'+2 f+\ell(\ell+1)\right)h_{rr} +\nonumber\\
&+4 r^2 (2 f+\ell(\ell+1)) h_{vv}+8 f r^3 h_{vv}'\, ,\label{eq:polar1}\\
 4i! r^4\left(\delta G\right)_{\ell\, vr} =&-2 f^2 r^3 h_{rr}'+\left(-2 r f'-\ell(\ell+1)+2\right)h_+ -f r^2 \left(4 r f'+2 f+\ell(\ell+1)\right)h_{rr} +\nonumber\\
&-2 r^2  \left(4 r f'+4 f+\ell(\ell+1)\right)h_{vr}+r \left(r f'-2 f\right) h_+'-8 f r^3 h_{vr}'+2 f r^2 h_+''+\nonumber\\
&-8 r^3 h_{vv}'-8 r^2 h_{vv}\, ,\label{eq:polar2}\\
 2 i! r^4\left(\delta G\right)_{\ell\, rr} =&r^2\left(2 r f'+\ell(\ell+1)\right) h_{rr} +2 f r^3 h_{rr}'+8 r^3 h_{vr}'-2 r^2 h_+''+4 r h_+'-4 h_+\, ,\label{eq:polar3}\\
2 i! \left(\delta G\right)_{\ell\, vu} =&h_{vr} f'+f h_{vr}'+2 h_{vv}'\, ,\label{eq:polar4}\\
4 i! r^3\left(\delta G\right)_{\ell\, ru} =&r^2 \left(r f'+2 f\right)h_{rr} -4 r^3 h_{vr}'+8 r^2 h_{vr}-2 r h_+'+4 h_+\, , \\
4i! r^2\left(\delta G\right)_{\ell\, +} =&-4 r^2 \left(4 r f'+4 f+\ell(\ell+1)-4\right)h_{vr} -f r^3 \left(r f'+2 f\right) h_{rr}'-4 r^3 \left(r f'+2 f\right) h_{vr}'+\nonumber\\
&+2 r \left(r f'-2 f\right)h_+' +4 \left(f-r f'\right) h_+-r^2 \left(4 f^2+f \left(6 r f'+\ell(\ell+1)-4\right)+r^2 f'^2\right)h_{rr} +\nonumber\\
&+2 f r^2 h_+''-8 r^4 h_{vv}''-16 r^3 h_{vv}'\, ,\label{eq:polar5}\\
4i! \left(\delta G\right)_{\ell\, -} =& f h_{rr}+4 h_{vr}\, .\label{eq:polar6}
}

\subsubsection{Sources}Below we list the nonzero components of sources for Einstein equations. 
Sources for the $i$-th order perturbation equations can be found in the following way (see e.g.\ appendix A of \citep{R19}). Let's assume that we already know the solution to perturbation Einstein equations up to the $i$-th order (it consists of metric perturbations $\pert{j} h_{\mu\nu}$ with $j\leq i$):
\eq{
\tilde{g}_{\mu\nu}=\bar g_{\mu\nu}+\sum\limits_{j=1}^{i}\sum\limits_{\ell} {\pert{j}{h_\ell}}_{\mu\nu}   \frac{a^j}{j!}    \, .
}
Using this solution we can calculate the Einstein tensor $G_{\mu\nu}(\tilde g)$. Although this tensor vanishes up to the order $i$, it contributes to the $i+1$ (and higher) perturbation equations. Finally, the source of the order $i+1$ is given by:
\eq{
{\pert{i+1}{S}}_{\mu\nu}&=[i+1]\left(-G_{\mu\nu}(\tilde{g})\right)\, ,
}
where $[k]\left(...\right)$ denotes the $k$-th order expansion of a given quantity. Although in most cases expressions for the sources ${\pert{i+1}{S}}_{\mu\nu}$ are complicated, their construction is a purely algebraic task and can be easily performed using computer algebra. Below we list nonzero components of $i$-th order sources in terms of explicit solutions ${}^{(j)}h_{\mu\nu}$ found for lower orders.

The source for the second order:
\eq{
{}^{(2)}S_{0\, vv}&=4 \left(1-\frac{r^2}{4M^2}\right)\Omega_{11}^2  \, ,\label{eq:s201}\\
{}^{(2)}S_{0\, vr}&=  -8 \Omega_{11}^2 \, ,\\
{}^{(2)}S_{0\, +}&=  -16 \Omega_{11}^2 \, ,\label{eq:s202}\\
{}^{(2)}S_{2\, vv}&= \left(\frac{r^2}{M^2}-8\right) \Omega _{11}^2 \, ,\label{eq:s221}\\
{}^{(2)}S_{2\, vr}&= 8 \Omega _{11}^2\, ,\\
{}^{(2)}S_{2\, vu}&=  \frac{8}{3} r \Omega _{11}^2\, ,\\
{}^{(2)}S_{2\, +}&=    16 r^2 \Omega _{11}^2  \, .\label{eq:s222}
}

The sources for the third order are zero.

\section{Gauge transformations}\label{appc}

Consider a gauge transformation induced by a gauge vector $\xi = \sum\limits_{i=0}^\infty \frac{a^i}{i!}{}^{(i)}\xi$. According to \citep{B97}, metric perturbations transform in the following way:
\eq{
\pert{1}h_{\mu\nu}\rightarrow\pert{1}h_{\mu\nu} &+\Lxi{1} \bar g _{\mu\nu}\, ,\label{gg1}\\
\pert{2}h_{\mu\nu}\rightarrow\pert{2}h_{\mu\nu} &+(\Lxi{2}+\Lxi{1}^2)\bar g_{\mu\nu} + 2\Lxi{1}\pert{1}h_{\mu\nu}\, ,\label{gg2}\\
\pert{3}h_{\mu\nu}\rightarrow\pert{3}h_{\mu\nu} &+(\Lxi{1}^3+3\Lxi{1}\Lxi{2}+\Lxi{3})\bar g_{\mu\nu} + 3(\Lxi{1}^2+\Lxi{2})\pert{1}h_{\mu\nu}+3\Lxi{1}\pert{2}h_{\mu\nu}\, .\label{gg3}
}

An explicit form of \eqref{gg1}-\eqref{gg3} for a gauge vector of order $i$ acting on a metric components of order $i$ reads (for clarity, we omit $i$ indices, dots and primes correspond to derivatives with respect to $v$ and $r$, respectively):
\eq{
 h_{\ell\, v\varphi}\rightarrow & h_{\ell\, v\varphi}-\dot{\xi}_\varphi\, ,\label{ga1}\\
 h_{\ell\, r\varphi}\rightarrow & h_{\ell\, r\varphi}+\frac{2  \xi _\varphi}{r}-  \xi  _\varphi'\, ,\label{ga2}\\
 h_{\ell\, u\varphi}\rightarrow & h_{\ell\, u\varphi}+  \xi  _\varphi\, ,\label{ga3}\\
 h_{\ell\, vv}\rightarrow&  h_{\ell\, vv}-\frac{1}{4} \left(f   \xi _r +2  \xi _v \right) f' +2\dot{\xi}_{v}\, ,\label{gp1}\\
 h_{\ell\, vr}\rightarrow & h_{\ell\, vr}+\frac{1}{2}f'   \xi  _r  + \xi _v' + \dot{\xi}_{r}\, ,\label{gp2}\\
 h_{\ell\, rr}\rightarrow & h_{\ell\, rr}+2  \xi _r' \, ,\label{gp3}\\
 h_{\ell\, +}\rightarrow & h_{\ell\, +}+2 r f   \xi _r -\ell (\ell+1)  \xi _u +4 r  \xi _v \, ,\label{gp4}\\
 h_{\ell\, -}\rightarrow & h_{\ell\, -}- \xi _u\, ,\label{gp5}\\
 h_{\ell\, vu}\rightarrow & h_{\ell\, vu} -\xi_v-\dot{\xi}_u\, ,\label{gp7}\\
  h_{\ell\, ru}\rightarrow & h_{\ell\, ru} - \xi_r+\frac{2}{r}\xi_u-\xi_u' \, .\label{gp6}
}


\bibliography{references}
\bibliographystyle{apsrev4-1}

\end{document}